\documentclass[runningheads]{llncs}
\usepackage[colorinlistoftodos]{todonotes}
\usepackage{amsmath}
\usepackage[T1]{fontenc}
\usepackage{float}
\usepackage{tabularx}
\usepackage[table]{xcolor}

\usepackage{graphicx}
\usepackage{subcaption}

\begin{document}
\title{Comparative Analysis of GAN and Diffusion for MRI-to-CT translation}
%
%

\author{Emily Honey\inst{1} \and Anders Helbo\inst{1} \and Jens Petersen\inst{1,2}}

\institute{Department of Computer Science, University of Copenhagen, Copenhagen, Denmark \and Department of Oncology, Rigshospitalet, Copenhagen, Denmark}

%
\maketitle              
\begin{abstract} Computed tomography (CT) is essential for treatment and diagnostics; In case CT are missing or otherwise difficult to obtain, methods for generating synthetic CT (sCT) images from magnetic resonance imaging (MRI) images are sought after. Therefore, it is valuable to establish a reference for what strategies are most effective for MRI-to-CT translation. In this paper, we compare the performance of two frequently used architectures for MRI-to-CT translation: a conditional generative adversarial network (cGAN) and a conditional denoising diffusion probabilistic model (cDDPM). We chose well-established implementations to represent each architecture: Pix2Pix for cGAN, and Palette for cDDPM. We separate the classical 3D translation problem into a sequence of 2D translations on the transverse plane, to investigate the viability of a strategy that reduces the computational cost. We also investigate the impact of conditioning the generative process on a single MRI image/slice and on multiple MRI slices. The performance is assessed using a thorough evaluation protocol, including a novel slice-wise metric Similarity Of Slices (SIMOS), which measures the continuity between transverse slices when compiling the sCTs into 3D format. Our comparative analysis revealed that MRI-to-CT generative models benefit from multi-channel conditional input and using cDDPM as an architecture.

\end{abstract}
%
%

\section{Introduction}
Generation of synthetic CTs (sCTs) from MRI images has multiple possible benefits \cite{doi:10.1148/radiol.2020204045}, for instance, as a source of electron density information for radiotherapy planning without the complexities of an additional CT scan. 

Several approaches have been suggested for solving this task. Generally, it has been observed that supervised methods relying on paired images achieved better results than unsupervised methods \cite{DAYARATHNA2024103046}. Many models for MRI-to-CT are based on generative adversarial networks (GANs) such as the 3D cycle-GAN model, which, according to Roberts M. et al \cite{roberts2023imaging}, were able to produce satisfactory sCTs. In a supervised setting, conditional GANs (cGANs) or more specifically cycle-GAN architectures are often extended to improve the results of MRI-to-CT translation. Examples of such extensions included the incorporation of spatial attention \cite{emami2020attention} and conditioning the generator on three MRI slices \cite{tie2020pseudo}.

Diffusion models, such as conditional versions of the Denoising Diffusion Probabilistic Model (cDDPM), have been deployed for CT synthesis as well \cite{lyu2022conversionMEDDIF,li2023ddmm}. Dayarathna S. et al \cite{DAYARATHNA2024103046} conclude that cDDPMs perform better than cGANs when synthesising the brain, whereas cGANs outperform cDDPMs in the pelvic region. 

Pix2Pix \cite{isola2017image-PIX2PIX}, a cGAN, and Palette, a cDDPM \cite{saharia2022palette-PALETTE} have performed well across multiple domains and tasks \cite{saharia2022palette-PALETTE,isola2017image-PIX2PIX}. This flexibility suggests reasonable results when applied to MRI-to-CT translation. Though comparisons of models for MRI-to-CT translation have been done in the development of new models, few independent comparisons have been made. 

This article offers an in-depth, unbiased comparison of two well-known architectures implementing a cGAN and a cDDPM. The basis for the experimentation is the well-known image-to-image (I2I) translation models Pix2Pix \cite{isola2017image-PIX2PIX} and Palette \cite{saharia2022palette-PALETTE}, using the publicly available implementations \cite{palette_implementation,CycleGANpix2pix}. All our code is available at \url{https://github.com/AHelbo/MRI2CT}. Translation in 2D lowers the computational cost and model complexity and enables easier parallel processing of 3D volumes. These benefits make 2D translation advantageous for MRI-to-CT applications, but make the synthetic data prone to issues with discontinuity. Our analysis accounts for this as the quality of resampling is measured through segmentation in 3D, and a novel slice-to-slice continuity metric the SIMOS. 
\section{Background}
A GAN is a system of two networks: The generator, G, and the discriminator, D, which are trained adversarially \cite{goodfellow2014generative-GAN}. GANs produce images from random noise \cite{isola2017image-PIX2PIX}, but cGANs are provided with an additional conditional input that influences and guides the generative process \cite{pang2021image-METHODS}. 

The objective function of Pix2Pix includes a traditional $\mathcal{L}_1\text{-loss}$. Previous work on cGANs has shown that adding such losses, $\mathcal{L}_1$ or $\mathcal{L}_2$, was beneficial for capturing low frequencies in the synthetic output \cite{isola2017image-PIX2PIX}. Furthermore, the writers claim that since the $\mathcal{L}_1\text{-loss}$ penalizes low-frequency errors in the generated images, it incentivizes modelling a discriminator that focuses on the high frequencies. This led to the PatchGAN discriminator, which classifies images as synthetic or real on $N \times N$ patches across the entire image before averaging the local results to determine the authenticity of the entire image \cite{isola2017image-PIX2PIX}.


Diffusion models are a class of generative networks, consisting of two stages: a forward diffusion stage and a backward denoising process \cite{croitoru2023diffusionSURVEY}.

In cDDPMs, both stages are Markov chains. The forward process gradually adds noise in T steps from $y_0$, a noise-free image, up to $y_T$, an image indistinguishable from Gaussian noise. It is possible to arbitrarily sample a noisy image, $y_t$, at any given noise level $t$ in DDPMs \cite{ho2020denoisingDDPM}. The backward process is finite and fixed to exactly $T$ steps and reverses the forward process by performing denoising steps so the synthetic image increasingly imitates the target distribution \cite{ho2020denoisingDDPM}.

An image pair $(x, y)$, where $y$ is conditioned on $x$, and a noise level $t \in [0; T]$ is sampled during training. Gaussian noise dependent on $t$ is then added to $y$ before a gradient descent step is taken based on the model's ability to predict the amount of added noise. Prediction is performed by $f_\theta$ a neural network, typically a U-Net \cite{saharia2022palette-PALETTE,dhariwal2021diffusionBEATS,superResolution}.

Inference in a cDDPM generates images through the backward process, where $f_\theta$ at each noise level from T to 0 predicts all the noise conditioned on the input image. Starting from $t=T$, noise is removed to denoise the synthetic image to noise level $t-1$ iteratively until $t=0$.

Palette concatenates the conditional image and the denoised image in each iteration, an approach inspired by previous work by Saharia \cite{saharia2022palette-PALETTE,superResolution}. In training, a noise schedule of $(1e^{-6}, 0.01)$ is applied in 2000 time-steps, and during inference, they have 1000 time-steps and a linear noise schedule of $(1e^{-4}, 0.09)$ \cite{saharia2022palette-PALETTE}.
\subsection{Related work}

GAN-based implementations applied to MRI-to-CT synthesis were developed as early as 2018 \cite{nie2018medical}. Nie et al. (2018) designed a cGAN network and experimented with its performance on medical I2I translation tasks, MRI-to-CT, and 3T-to-7T. Notably, the generator synthesised overlapping source image patches and fused them into a single output image by averaging the overlapping regions \cite{nie2018medical}. The model was applied to two separate datasets containing brain and pelvic scans. The authors concluded that their proposed method outperformed other I2I methods across datasets by achieving better Peak Signal-to-Noise ratio (PSNR) and Mean Absolute Error (MAE) scores \cite{nie2018medical}.

In 2019, experimentation and comparison of U-Net and GAN-based models for MRI-to-CT translation was performed \cite{kaiser2019MriCtGans}, one of which was directly based on the architecture of Pix2Pix \cite{isola2017image-PIX2PIX}. Two out of three models used a U-Net, both of which solved the task in 2D; the last model was a context-aware GAN for medical 3D I2I translation presented by Nie et al. \cite{nie2017medical}. Importantly, their work showcased the positive impact of the adversarial element in the GAN architecture instead of solely relying on U-Nets for image generation in medical I2I tasks \cite{kaiser2019MriCtGans}.
Diffusion models have also been developed for MRI-to-CT translation. Lyu and Wang \cite{lyu2022conversionMEDDIF} employ four strategies for diffusion and compare their performance to a CNN and a GAN-based solution. The DDPM achieves the highest Structural Similarity Index Measure (SSIM) and PSNR \cite{lyu2022conversionMEDDIF} though their GAN implementation tends to hallucinate 'severe' artefacts in sCT, whereas the diffusion models do not exhibit this behaviour \cite{lyu2022conversionMEDDIF}.

A substantial amount of the existing work on MRI-to-CT translation uses 3D architectures \cite{nie2018medical,kaiser2019MriCtGans,roberts2023imaging}, which has a significantly higher computational cost \cite{kaiser2019MriCtGans}.

Several sources \cite{nie2018medical,kaiser2019MriCtGans} conclude that adversarial learning guides the generative process in a positive direction. The results from \cite{lyu2022conversionMEDDIF} indicate that diffusion models are more suitable for the task, even without the advantage of a discriminator.
\section{Method}

\subsection{Evaluation}

We evaluate the sCTs on Mean Squared Error (MSE), MAE, PSNR, SSIM, Fréchet inception distance (FID), SIMOS and a segmentation-based intersection over union (IoU) metric. There are advantages and disadvantages to all metrics, but in combination, they provide a good insight into the performance of our models.

The MAE, MSE and PSNR are pixel-wise metrics that measure the accuracy at a pixel level. This makes them more computationally efficient compared to the SSIM, which captures perceptual and structural differences. The SSIM discriminates structural changes between the synthetic and target images. In medical images, 'structures' are shapes such as bones, soft tissue, and body outlines. 

FID measure the distance between the distribution of two domains, in this context, the domains are the synthetic and the target data. While FID cannot detect overfitting \cite{lucic2018gansFIDGANS}, Heusel et al. \cite{heusel2017gansFID} claim to see a correlation between the FID and human judgment, which makes it a valuable measure for evaluation during experimentation.

\subsubsection{Similarity Of Slices.}\label{SIMOS}

Due to the potential problems of solving a 3D task as a sequence of 2D tasks, we developed a metric that measures 3D continuity across resampled slices. We define $\text{SIMOS}(y, \Tilde{y})$ as: 

\begin{equation}\label{eq:simos}
    \text{SIMOS}(y, \widetilde{y}) = \frac{1}{N-1} \sum_{i = 0}^{N-1} | \text{MSE}(y_{i}, y_{i + 1}) - \text{MSE}(\widetilde{y}_i, \widetilde{y}_{i + 1})|
\end{equation}

Where $y$ is the ground truth image and $\widetilde{y}$ is the synthetic image. SIMOS is given by the MAE of the accumulated difference in the MSE from consecutive slices in the input images. A small value correlates to a small difference between slice pairs. If $y = \Tilde{y}$ SIMOS will be zero. 

\subsubsection{Segmentation.}
We use an image segmentation method to display the model's ability to correctly synthesize different tissues, sizes, and positions. To balance computational demands and processing time, segmentation was only performed on 50\% of the test set.
The sCT slices are resampled to Nifti format before performing segmentation. We utilize TotalSegmentator \cite{total_segmentator} for segmenting in 3D, and the Segment Anything Model (SAM) \cite{segmentanything2023} to segment in 2D. For both 2D and 3D segmentation, we subsequently calculate the mean IoU between each ground truth mask and the corresponding synthetic mask.

\begin{figure}[h]
    \centering
    \begin{subfigure}{0.25\textwidth}
        \centering
        \includegraphics[width=\linewidth]{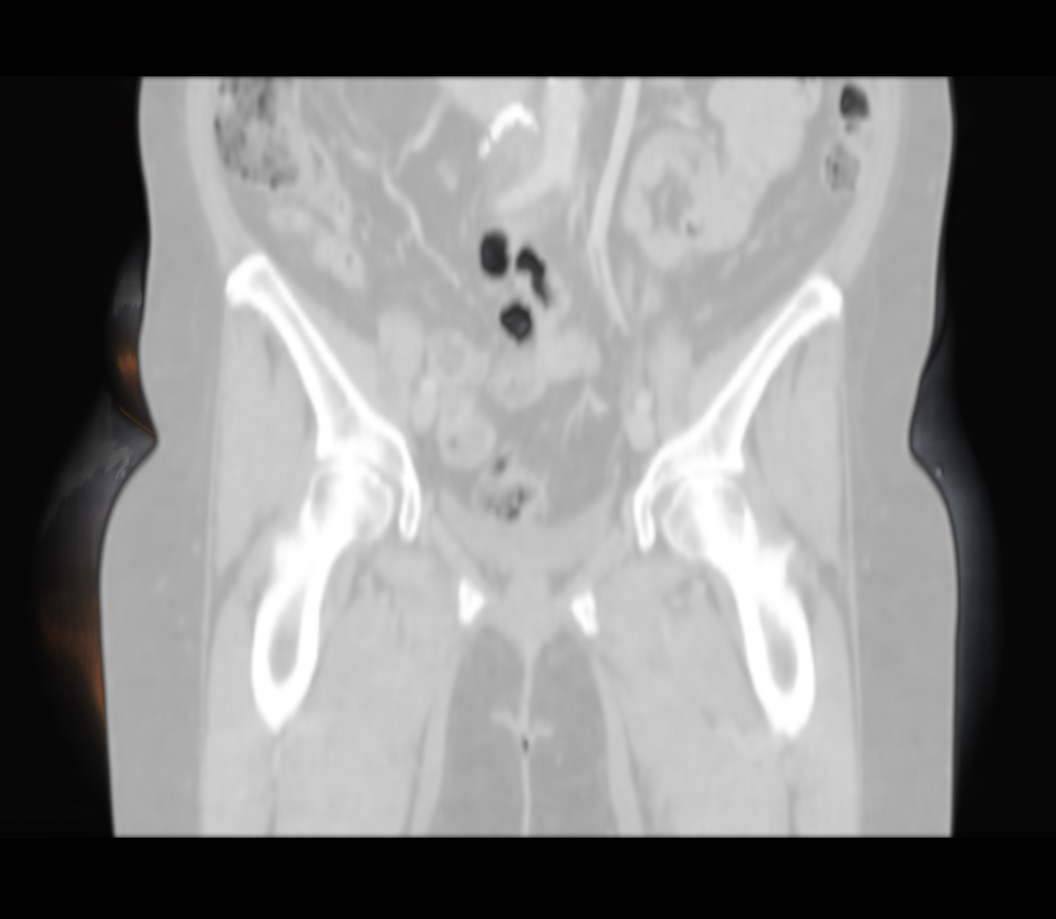} 
        \begin{minipage}{0.9\linewidth} 
            \caption{GT}
        \end{minipage}
    \end{subfigure}%
    \hfill 
    \begin{subfigure}{0.25\textwidth}
        \centering
        \includegraphics[width=\linewidth]{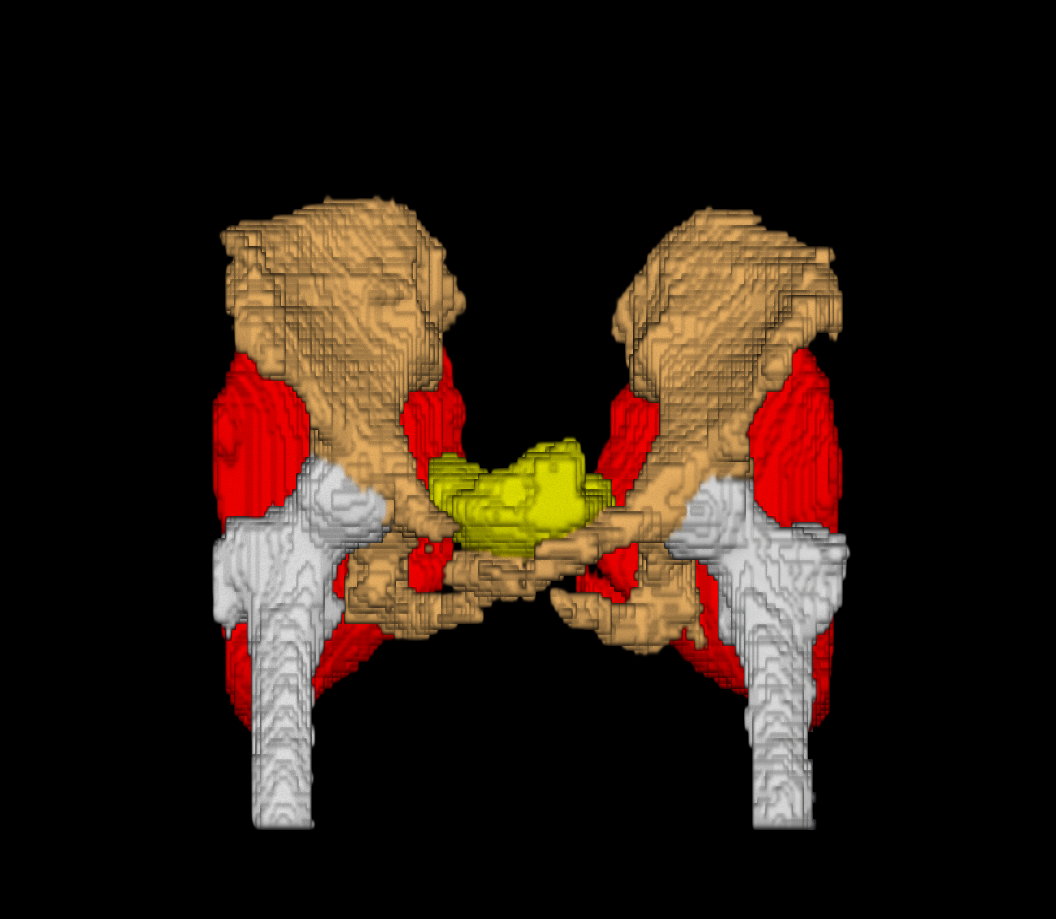}
        \begin{minipage}{0.9\linewidth}
            \caption{Segmented GT}            
        \end{minipage}
    \end{subfigure}%
    \hfill
    \hfill 
    \begin{subfigure}{0.25\textwidth}
        \centering
        \includegraphics[width=\linewidth]{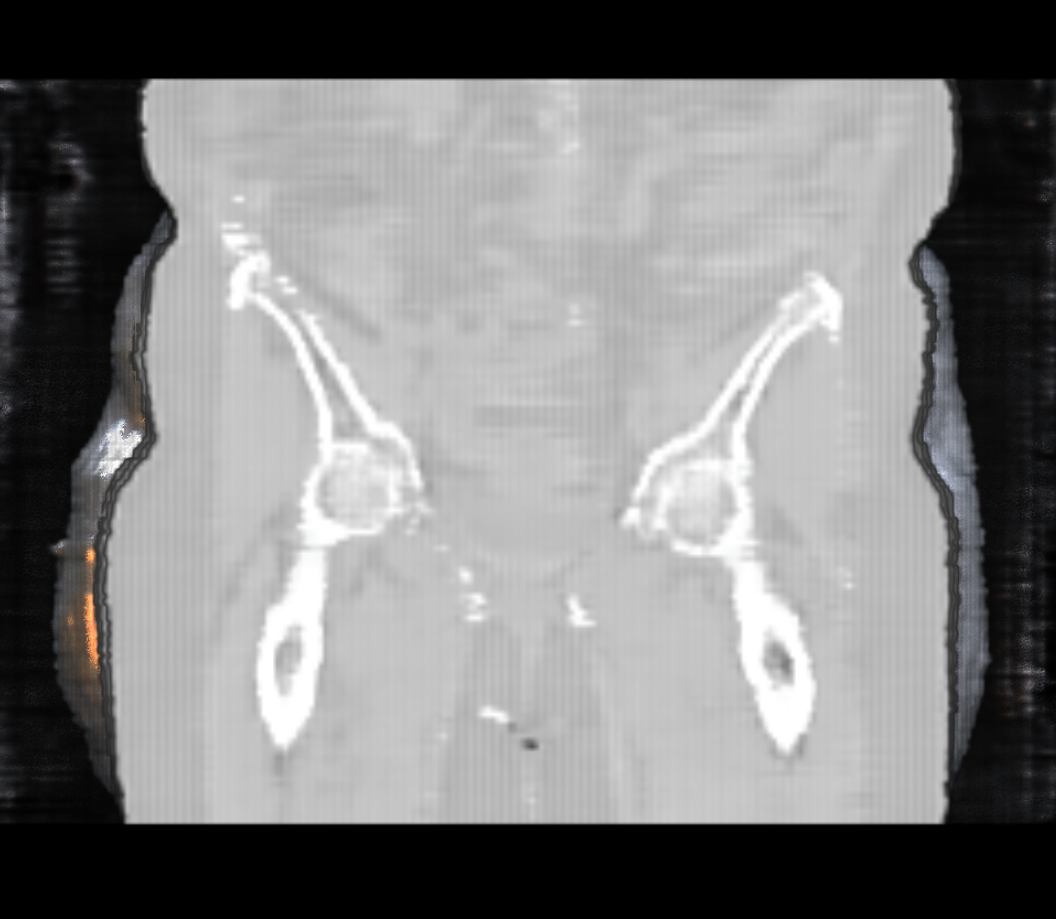}
        \begin{minipage}{0.9\linewidth}
            \caption{sCT}        
        \end{minipage}
    \end{subfigure}%
    \hfill    
    \begin{subfigure}{0.25\textwidth}
        \centering
        \includegraphics[width=\linewidth]{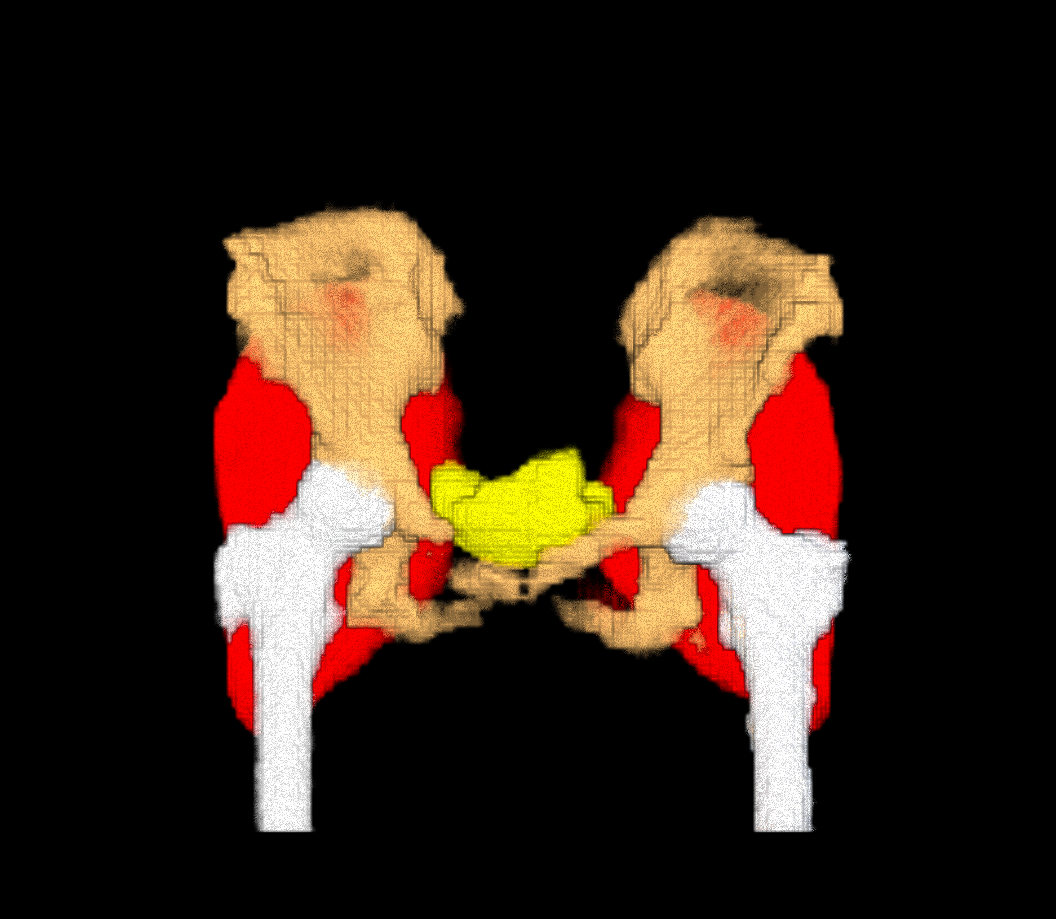}
        \begin{minipage}{0.9\linewidth}
            \caption{Segmented sCT}            
        \end{minipage}
    \end{subfigure}
    \caption{Example of 3D segmentation using TotalSegmentator on the same patient. One can see the segmentations masks of the femurs (white), gluteus maximus (red), the urinary bladder (yellow) and the hipbones (orange).}
    \label{fig:segmented3D}
\end{figure}

\subsection{Data}\label{data}

The dataset is sourced from the SynthRAD2023 Grand Challenge \cite{challengeDATA}, it contains paired brain and pelvic MRI/CT scans in NIfTI format, collected from 360 patients across three Dutch hospitals. The goal of SynthRAD2023 was to enable comparison of methods for sCT generation from MRI images, and the data have already been preprocessed for this purpose \cite{SynthRAD2023_preprocessing}. We split the data set on a per-patient level into a training, validation, and test set, each with an even distribution of brain/pelvic scans and hospitals.

\subsubsection{Preprocessing.} The dataset is compiled into a sequence of 2D slices aligned on the transverse plane across modalities. Each 2D slice is used as a data point. 

\subsubsection{CT-specific preprocessing.} 
\begin{figure}[H]
    \centering
    \begin{minipage}{0.45\textwidth}
        \centering
        \includegraphics[width=\linewidth]{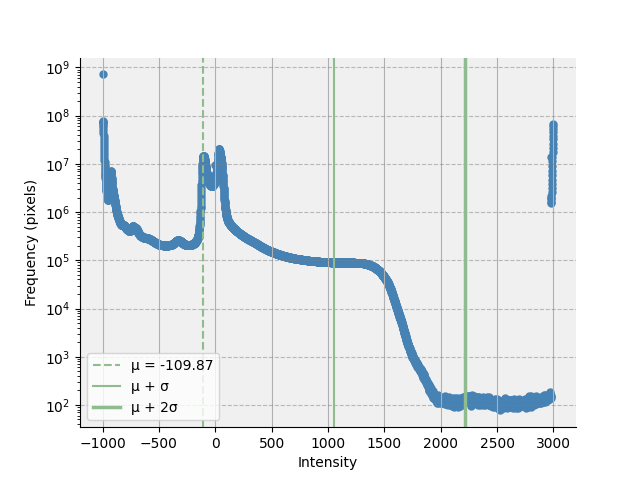}
        \caption{Frequency plot for the unprocessed CT scans}
        \label{CTraw}
    \end{minipage}
    \begin{minipage}{0.45\textwidth}
        \centering
        \includegraphics[width=\linewidth]{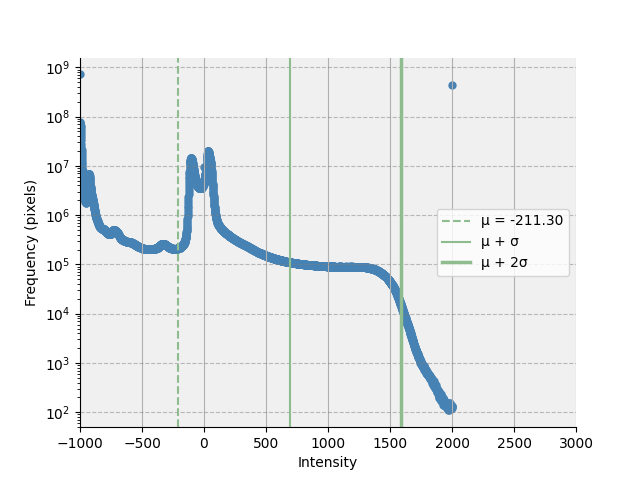}
        \caption{Frequency plot of the CT data after processing}
        \label{CTprocessed}
    \end{minipage}
\end{figure}

Initially, values ranged between $[-1000; 3000]$ HU. Values outside the range $[-1000;2000]$ are likely abnormalities such as metal implants. The intensities are therefore capped to the upper limit of $2000$ HU. A min-max-normalization is applied with the population minimum value, $-1000$ HU, and the upper limit as a maximum value. The normalized values are then mapped into the range $[0; 1]$. A plot of the frequency distribution of voxel intensities before and after preprocessing is provided in Fig. \ref{CTraw} and Fig. \ref{CTprocessed}.

\subsubsection{MRI-specific preprocessing.} 
\begin{figure}[H]
    \centering
    \begin{minipage}{0.45\textwidth}
        \centering
        \includegraphics[width=\linewidth]{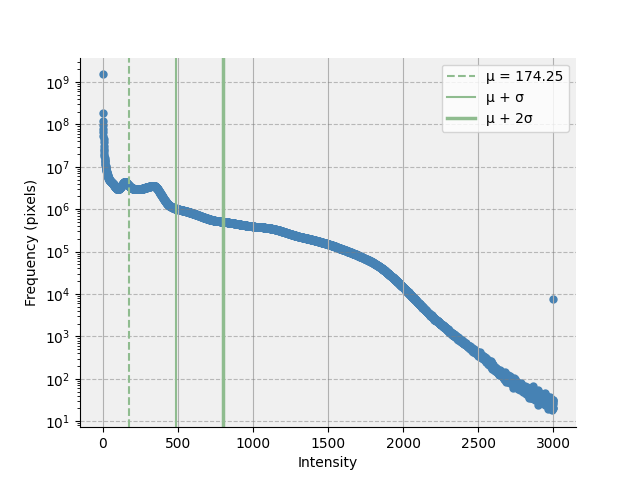}
        \caption{Frequency plot for the unprocessed MRI scans}
        \label{MRIraw}
    \end{minipage}
    \begin{minipage}{0.45\textwidth}
        \centering
        \includegraphics[width=\linewidth]{Images/Preprocessing/CT-cap2000.png}
        \caption{Frequency plot of the MRI data after processing}
        \label{MRIProcessed}
    \end{minipage}
\end{figure}

The frequency distribution of voxel intensities features a long tail of infrequent intensities, as the distribution of MRI voxel intensities varies significantly due to differences in hardware and imaging process \cite{mr_signal}. To mitigate the influence of extreme values while preserving the relative intensity distribution, intensities beyond the 98th percentile are capped at the 98th percentile value locally for each image.
A plot of the frequency distribution of the voxel intensities before and after the preprocessing step is provided in Fig. \ref{MRIraw} and Fig. \ref{MRIProcessed}.

\subsubsection{Multi-channel MRI.}\label{multi-channel-images} To investigate a computationally efficient way to leverage 3D information, a separate multi-channel dataset was constructed. MRI slices with preceding and subsequent slices were compiled into a single image. These were paired with the target CT slice corresponding to the middle MRI slice.

\section{Experiments}

\subsection{Experiment design}

The aim is to produce four models: $\text{cGAN}_{1}$, $\text{cGAN}_{3}$, $\text{cDDPM}_{1}$ and $\text{cDDPM}_{3}$. Each architecture is conditioned on both single-channel MRI and multi-channel MRI. Experimentation consists of three phases: hyperparameter fitting, model selection, and model evaluation. 

During the hyperparameter fitting phase, hyperparameters are chosen based on the SSIM, the PSNR and the training loss. The hyperparameter fitting phase is only performed on the single-channel models. The multi-channel models are configured with the same hyperparameters as the single-channel models. An overview of the tested hyperparameters for the cGAN- and cDDPM models can be seen in Table \ref{gan_hyper} and \ref{fig:diff_hyper}, respectively. This phase uses the training and validation sets. Plots of the training loss and metrics are available at \url{https://github.com/AHelbo/MRI2CT}.

In the model selection phase, all models are trained with the optimal hyperparameters. For each epoch, we calculate the PSNR, SSIM, FID \cite{PYTORCH_FID}, and SIMOS on the validation set. We select the epoch at which the model demonstrates optimal performance for the evaluation phase, where we deploy all previously used metrics and segmentation on the test set.

\subsection{cGAN - Hyperparameter fitting phase}

\begin{table}[h]
  \centering
    \begin{tabularx}{\textwidth}{|>{\hsize=0.3\hsize\arraybackslash}X|>{\hsize=0.4\hsize\arraybackslash}X|>{\hsize=0.3\hsize\arraybackslash}X|}
        \hline
        \textbf{Parameter} & \textbf{Tested values} & \textbf{Optimal value} \\
        \hline
        $\mathbf{\lambda}$\textbf{-value} & 50, 75, 100, 125, 150 & 100 \\ 
        \hline
        \textbf{Batch size} & 1, 5, 10  & 10 \\
        \hline
        \textbf{Learning rate} & $125e^{-5}$,$25e^{-5}$,$5e^{-5}$,$1e^{-4}$,$2e^{-4}$ & $5e^{-5}$ \\
        \hline
        $\mathbf{D_{freq}}$ & 1, 3, 5, 10, 20 & 10 \\ 
        \hline
    \end{tabularx}
    \caption{Summary of tested parameters for the cGAN-based models. $\lambda$-value is a multiplication factor that defines the weight of the loss function. $D_{freq}$ is the Discriminator frequency, which allows the discriminator to be updated at every n'th data point during training.}
    \label{gan_hyper}
\end{table}

\subsubsection{$\lambda$-value.} None of the values tested showed any notable improvements compared to the baseline $\lambda=100$.

\subsubsection{Batch normalization.} Models with batch sizes greater than 1 utilize batch normalization. Among the tested values, batch size 10 is preferred due to its superior score in $\mathcal{L}_1\text{-loss}$, SSIM and PSNR.

\subsubsection{Learning rate.} With learning rate $5e^{-5}$ the running loss exhibited fewer fluctuations, and the SSIM and PSNR metrics consistently improved and converged faster compared to lower learning rates. While higher learning rates eventually achieved similar SSIM and PSNR scores, they also introduced greater fluctuations in the running losses.

\subsubsection{Discriminator learning frequency.} Early experimentation revealed an unstable training, due to the discriminator becoming too good at distinguishing real and fake images too fast. This caused an imbalance between the networks, resulting in G receiving primarily negative feedback from D. Decreasing the frequency at which the discriminator is updated yielded a more stable and less volatile training. $D_{freq} = 10$ resulted in the most stable training and a higher SSIM than $D_{freq}=20$.
\subsection{cGAN - Model selection phase}

cGAN$_{1}$ and cGAN$_{3}$ performed best in epochs 625 and 685, respectively. The main criteria used to determine this were low SIMOS and FID scores. Later epochs had lower similarity between consecutive slices, as SIMOS decreased beyond this point, indicating a smaller discontinuity from one slice to the next. The FID showed a minimum in the selected and surrounding epochs. In the subsequent epochs, the FID increased, implying that the sCT images become more distinct from the ground truth images as training continues. Thus, the selected epoch was a compromise between SIMOS and FID. Furthermore, PSNR and SSIM also trend downwards after the selected epochs. Such a development in the PSNR and the SSIM on the validation set indicates overfitting.
\subsection{cDDPM - Hyperparameter fitting phase} 

\begin{table}[h]
  \centering
  \begin{tabularx}{\textwidth}{|>{\hsize=0.3\hsize\arraybackslash}X|>{\hsize=0.4\hsize\arraybackslash}X|>{\hsize=0.3\hsize\arraybackslash}X|}
    \hline
    \textbf{Parameter} & \textbf{Tested values} & \textbf{Optimal value} \\
    \hline
    
    \textbf{Learning rate} & $1e^{-4}$, $2e^{-4}$, $4e^{-4}$, $5e^{-5}$, $25e^{-5}$  &  $1e^{-4}$              \\
    \hline
    
    \textbf{Loss function} & $\mathcal{L}_1$, $\mathcal{L}_2$ & $\mathcal{L}_1$               \\ 
    \hline

  \end{tabularx}
    \caption{Summary of the hyperparameter fitting for the cDDPM-based models. The experiments are conducted in the order the parameters appear in the table.}
    \label{fig:diff_hyper}
\end{table}

\begin{figure}[h]
    \centering
    \begin{subfigure}{0.25\textwidth}
        \centering
        \includegraphics[width=\linewidth]{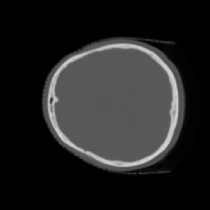} 
        \begin{minipage}{0.8\linewidth} 
            \caption{Successful brain sample}
        \end{minipage}
    \end{subfigure}%
    \hfill 
    \begin{subfigure}{0.25\textwidth}
        \centering
        \includegraphics[width=\linewidth]{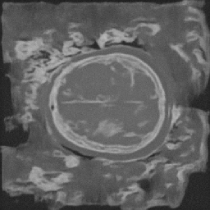}
        \begin{minipage}{0.8\linewidth}
            \caption{Failed brain sample}            
        \end{minipage}
    \end{subfigure}%
    \hfill
    \hfill 
    \begin{subfigure}{0.25\textwidth}
        \centering
        \includegraphics[width=\linewidth]{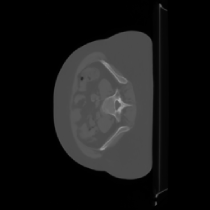}
        \begin{minipage}{0.8\linewidth}
            \caption{Successful pelvic sample}        
        \end{minipage}
    \end{subfigure}%
    \hfill    
    \begin{subfigure}{0.25\textwidth}
        \centering
        \includegraphics[width=\linewidth]{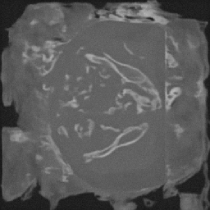}
        \begin{minipage}{0.8\linewidth}
            \caption{Failed pelvic sample}            
        \end{minipage}
    \end{subfigure}
    \caption{Illustration of the failed samples. The frequency of failed samples and the amount of noise in them decreased in the later epochs.}
    \label{fig:failed_pal}
\end{figure}

The cDDPM models frequently produced failed samples, i.e. samples with large amounts of noise and/or faint structures (see Fig. \ref{fig:failed_pal}) even when the model had generated higher-quality samples in the same or previous epochs. This impacted the score on our metrics, which led to us favouring hyperparameters that reduced the number of failed samples.

\subsubsection{Learning rate.} The experiments revealed no significant difference: the peak SSIM and PSNR across learning rates were approximately $0.8$ and $25$, and both the MSE and the MAE converged towards zero, albeit unstable. No tendencies were uncovered that could justify further experiments or choosing one learning rate over another; therefore, the learning rate was fixed to $1e^{-4}$ as done in \cite{saharia2022palette-PALETTE}.

\subsubsection{Loss functions.} We experimented with the $\mathcal{L}_1$ and $\mathcal{L}_2$ loss, the pixel-wise parameters, and visual inspection did not indicate that one was better, even though according to Saharia et al. \cite{saharia2022palette-PALETTE} $\mathcal{L}_1$ might reduce the number of potential hallucinations and yield a lower sample diversity. $\mathcal{L}_1$ lowered the occurrence of failed samples and caused the frequency of the failed samples to decrease faster and more steadily per epoch than $\mathcal{L}_2$. Therefore, $\mathcal{L}_1$ became the loss function for the cDDPM-based models.
\subsection{cDDPM - Model selection phase}

Sampling the sCTs from cDDPM models was limited by computational cost. Fewer failed samples appeared the more iterations the model had trained. We suspect that this trend is caused by models that have been trained for a shorter amount of time not being able to robustly map the latent space to within distribution samples. This could lead to a situation where the iterative predictive process of DDPMs, where output becomes input for the next iteration, moves samples further and further away from the sought distribution. 
This, combined with the poor running losses in the early epochs, indicated that an optimal epoch would not be achieved early in training. To work within the limitations, we sampled from epoch 200 onwards. We selected epoch 335 for cDDPM$_{1}$ and epoch 360 for cDDPM$_{3}$, since these models showcased the best scores across metrics and were thus the best available epochs.
\section{Results}
The selected models were evaluated on the test dataset. A summary of all results is presented in Table \ref{totalRes}. 
\newcommand{\GANSingleSSIM}{0.838}
\newcommand{\GANSinglePSNR}{25.978}
\newcommand{\GANSingleMSE}{206.882}
\newcommand{\GANSingleMAE}{12.305}
\newcommand{\GANSingleSIMOS}{53.953}
\newcommand{\GANSingleFID}{88.768}
\newcommand{\GANSingleSEGTwoD}{0.396}
\newcommand{\GANSingleSEGTreeD}{0.673}
\newcommand{\GANSingleTrainTime}{52.141}
\newcommand{\GANSingleSampleTime}{9.042}

\newcommand{\GANMultiSSIM}{0.841}
\newcommand{\GANMultiPSNR}{25.898}
\newcommand{\GANMultiMSE}{283.555}
\newcommand{\GANMultiMAE}{12.849}
\newcommand{\GANMultiSIMOS}{47.851}
\newcommand{\GANMultiFID}{93.579}
\newcommand{\GANMultiSEGTwoD}{0.394}
\newcommand{\GANMultiSEGTreeD}{0.59}
\newcommand{\GANMultiTrainTime}{51.02}
\newcommand{\GANMultiSampleTime}{9.126}

\newcommand{\DiffSingleSSIM}{0.872}
\newcommand{\DiffSinglePSNR}{26.194}
\newcommand{\DiffSingleMSE}{139.428}
\newcommand{\DiffSingleMAE}{3.387}
\newcommand{\DiffSingleSIMOS}{42.058}
\newcommand{\DiffSingleFID}{15.657}
\newcommand{\DiffSingleSEGTwoD}{0.584}
\newcommand{\DiffSingleSEGTreeD}{0.741}
\newcommand{\DiffSingleTrainTime}{4.682}
\newcommand{\DiffSingleSampleTime}{0.024}

\newcommand{\DiffMultiSSIM}{0.881}
\newcommand{\DiffMultiPSNR}{26.620}
\newcommand{\DiffMultiMSE}{114.616}
\newcommand{\DiffMultiMAE}{3.037}
\newcommand{\DiffMultiSIMOS}{22.968}
\newcommand{\DiffMultiFID}{14.152}
\newcommand{\DiffMultiSEGTwoD}{0.571}
\newcommand{\DiffMultiSEGTreeD}{0.717}
\newcommand{\DiffMultiTrainTime}{4.652}
\newcommand{\DiffMultiSampleTime}{0.024}

\begin{table}[h]
  \centering
  \begin{tabularx}{\textwidth}{|>{\centering\arraybackslash}l|>{\centering\arraybackslash}X|>{\centering\arraybackslash}X|>{\centering\arraybackslash}X|>{\centering\arraybackslash}X|}
    \hline
    \textbf{} & $\mathbf{\textbf{cGAN}_{1}}$ & $\mathbf{\textbf{cGAN}_{3}}$ & $\mathbf{\textbf{cDDPM}_{1}}$ & $\mathbf{\textbf{cDDPM}_{3}}$ \\
    \hline
    \textbf{SSIM ↑} & \GANSingleSSIM & \GANMultiSSIM & \DiffSingleSSIM & \textbf{\DiffMultiSSIM} \\
    \hline
    \textbf{PSNR ↑} & \GANSinglePSNR & \GANMultiPSNR & \DiffSinglePSNR & \textbf{\DiffMultiPSNR} \\
    \hline
    \textbf{Training (iters/sec) ↑} & \textbf{\GANSingleTrainTime} & \GANMultiTrainTime & \DiffSingleTrainTime & \DiffMultiTrainTime \\
    \hline
    \textbf{Sampling (samples/sec) ↑} & \GANSingleSampleTime & \textbf{\GANMultiSampleTime} & \DiffSingleSampleTime & \DiffMultiSampleTime \\
    \hline
    \textbf{2D segmentation IoU ↑} & \GANSingleSEGTwoD & \GANMultiSEGTwoD & \textbf{\DiffSingleSEGTwoD} & \DiffMultiSEGTwoD \\
    \hline

    \textbf{3D segmentation IoU ↑} & \GANSingleSEGTreeD & \GANMultiSEGTreeD & \textbf{\DiffSingleSEGTreeD} & \DiffMultiSEGTreeD \\
    \hline    
    \textbf{FID ↓} & \GANSingleFID & \GANMultiFID & \DiffSingleFID & \textbf{\DiffMultiFID} \\
    \hline
    \textbf{SIMOS ↓} & \GANSingleSIMOS & \GANMultiSIMOS & \DiffSingleSIMOS & \textbf{\DiffMultiSIMOS} \\
    \hline
  \end{tabularx}
  \caption{Summary of Image Processing Metrics by Model}
  \label{totalRes}
\end{table}

\subsubsection{SSIM and PSNR.}

The cDDPM models outperform the cGAN models on SSIM and PSNR. It is worth noticing that on these metrics the multi-channel models achieve higher scores than their single-layer counterparts, except for the PSNR of cGAN-models where cGAN$_{1}$ scores $25.978$ against $25.898$ for cGAN$_{3}$. The difference between the single-channel and multi-channel models is significantly higher for the cDDPM models. Judged on the PSNR, and SSIM cDDPM$_{3}$ is the best model.
\subsubsection{Time consumption.}

The time consumption is measured as the time required for training and sampling each model consecutively on the same GPU (NVIDIA GeForce GTX TITAN X). This approach allowed us to directly compare the time consumption of the models.

The selected cDDPM models were trained for $\sim 10$ days, whereas the GAN models required $\sim 3$ days. To get a more generalisable measure for the time consumption, we trained the models for $25,000$ iterations, the number of iterations is divided by the elapsed time. Similarly, to measure sample speed, we sampled $5,000$ sCT slices and divided the number of samples by the elapsed time (see Table \ref{totalRes}).

The cGAN-based models were the fastest in training and sampling time. On average, cGAN-based models accomplish approximately $11.05$ iterations a second more during training. Sample time revealed an even bigger difference, as the cGAN models, on average, sample data $378.5$ times faster than the cDDPM models. 
\subsubsection{Segmentation.}
The cDDPM-based models score higher than the cGAN-based models in 3D and 2D segmentation, with $\text{cDDPM}_{1}$ achieving the highest IoU scores of \text{\DiffSingleSEGTwoD} and \text{\DiffSingleSEGTreeD} for 2D and 3D, respectively. Noticeably, multi-channel conditional input seems to have no positive influence on the models on this aspect, as the multi-channel models perform worse than their single-channel peers.

\begin{figure}[h]
    \centering
    \begin{subfigure}{0.25\textwidth}
        \centering
        \includegraphics[width=\linewidth]{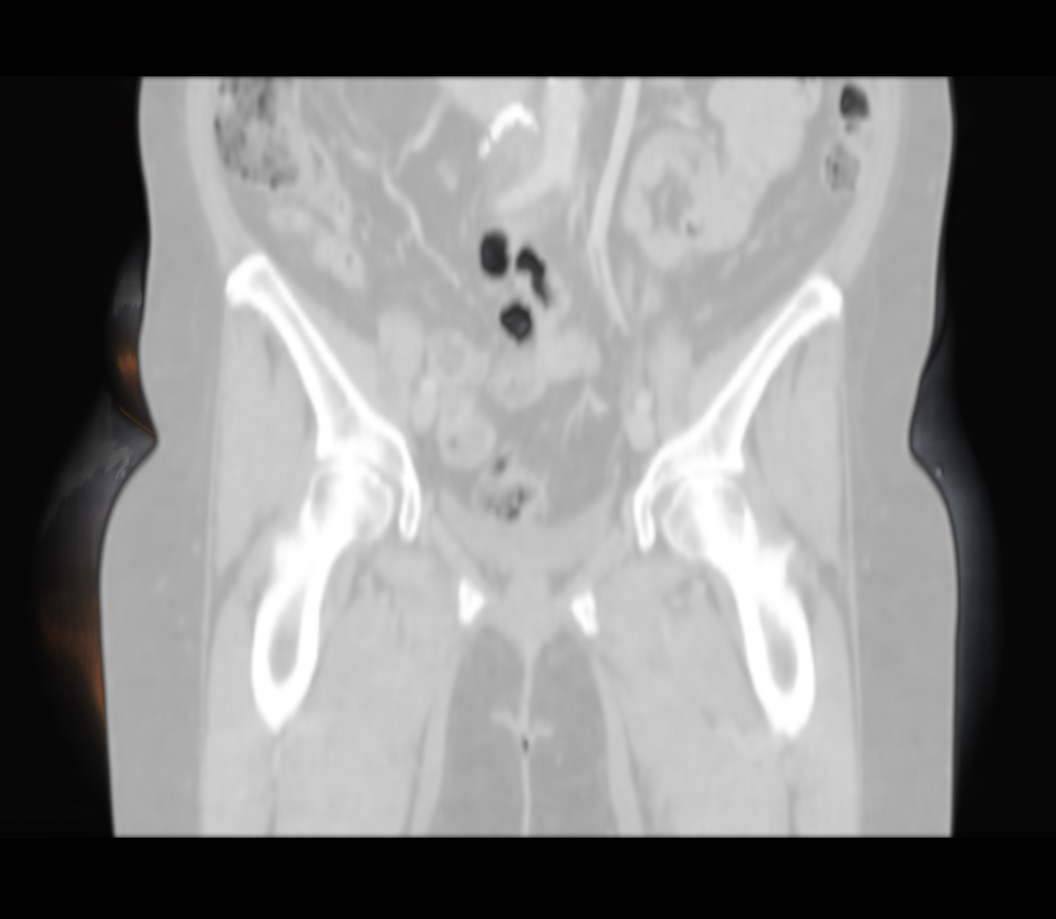} 
        \begin{minipage}{0.9\linewidth} 
            \caption{GT}
        \end{minipage}
    \end{subfigure}%
    \hfill 
    \begin{subfigure}{0.25\textwidth}
        \centering
        \includegraphics[width=\linewidth]{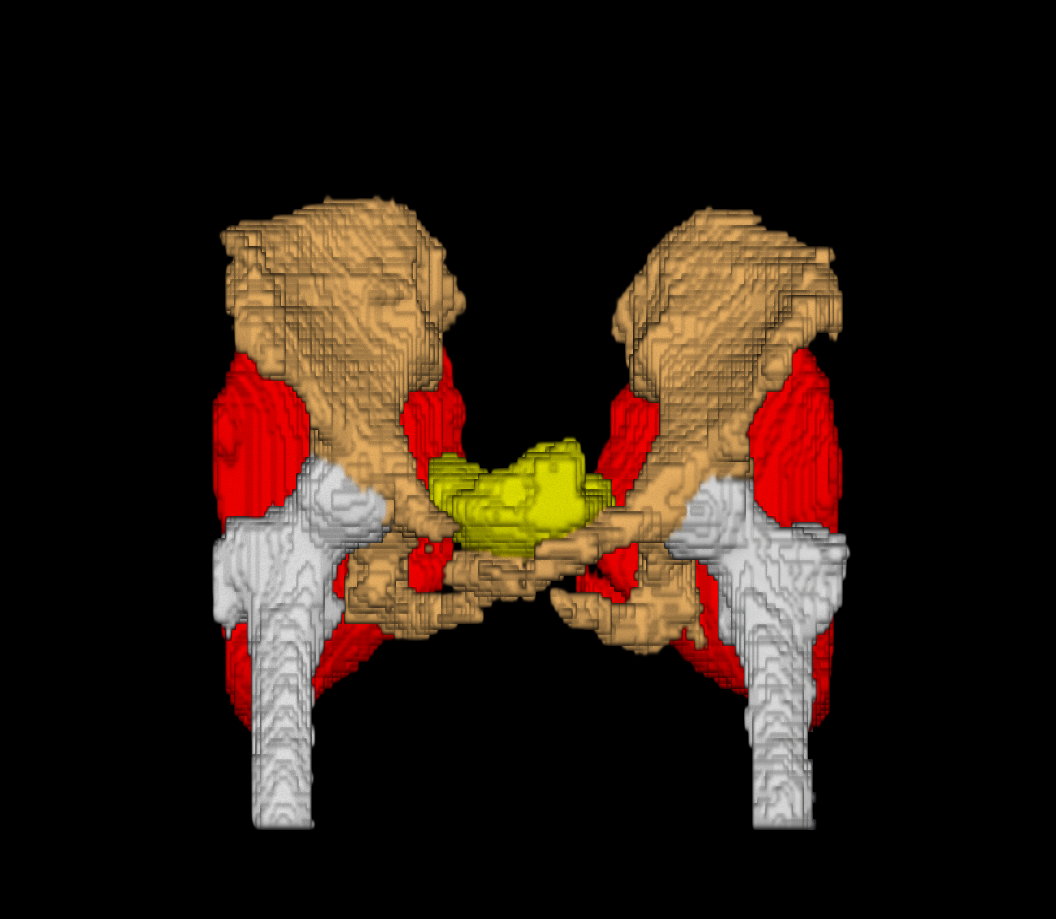}
        \begin{minipage}{0.9\linewidth}
            \caption{Segmented GT}            
        \end{minipage}
    \end{subfigure}%
    \hfill
    \hfill 
    \begin{subfigure}{0.25\textwidth}
        \centering
        \includegraphics[width=\linewidth]{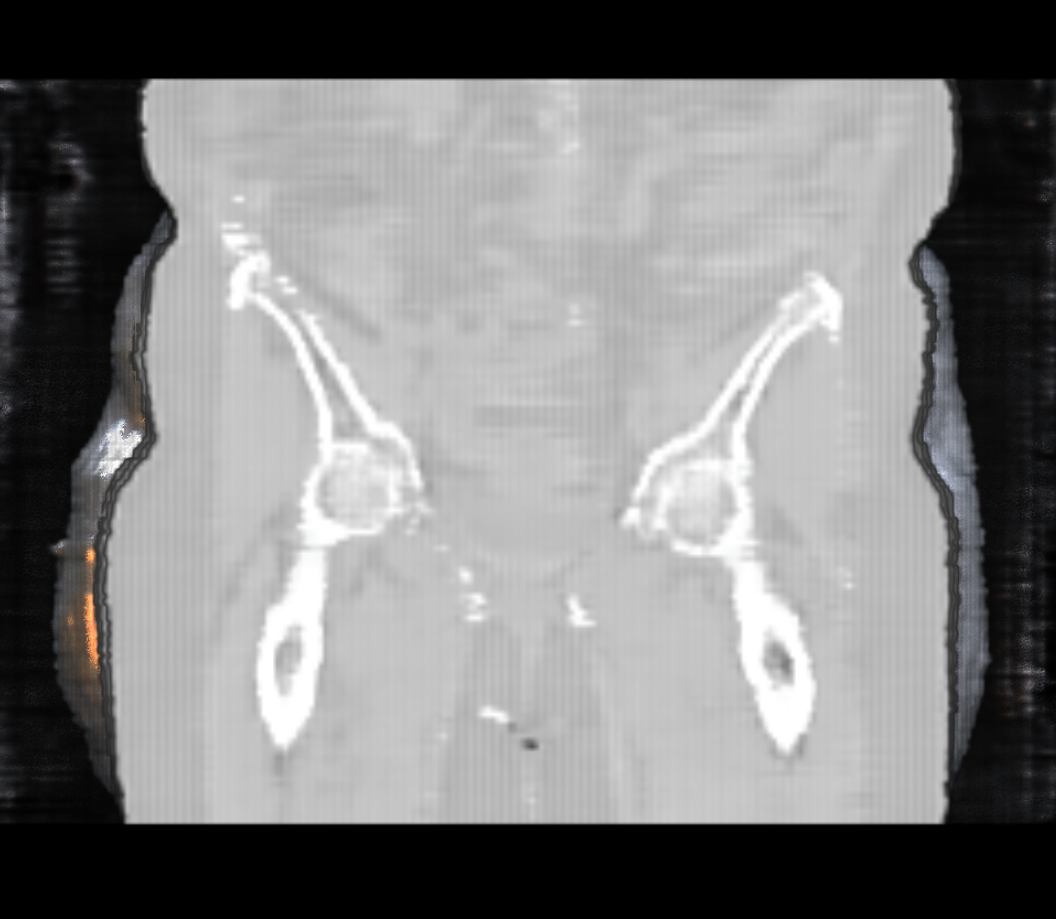}
        \begin{minipage}{0.9\linewidth}
            \caption{sCT}        
        \end{minipage}
    \end{subfigure}%
    \hfill    
    \begin{subfigure}{0.25\textwidth}
        \centering
        \includegraphics[width=\linewidth]{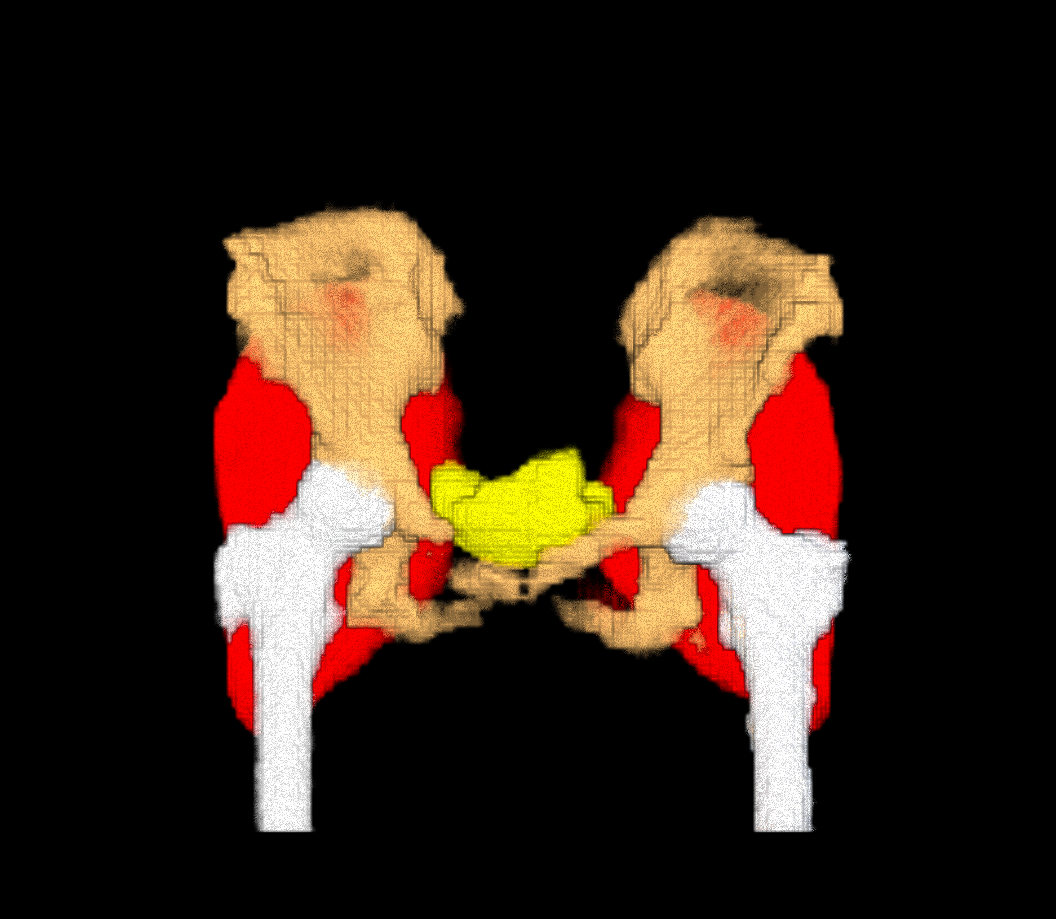}
        \begin{minipage}{0.9\linewidth}
            \caption{Segmented sCT}            
        \end{minipage}
    \end{subfigure}
    \caption{Example of 3D segmentation using TotalSegmentator on the same patient. One can see the segmentations masks of the femurs (white), gluteus maximus (red), the urinary bladder (yellow) and the hipbones (orange).}
    \label{fig:segmented3D}
\end{figure}

\subsubsection{FID.}
The cGAN models do not appear to benefit from supplying the generator with a multi-channel conditional input based on the FID score. The cDDPM models do however as we observe a lower FID score in cDDPM$_{3}$ than cDDPM$_{1}$. Generally, the cDDPM models perform significantly better than the cGAN models on the FID, with cDDPM$_{3}$ achieving the best overall FID.
\subsubsection{SIMOS.}
The SIMOS score of the cGAN and cDDPM models indicates that both architectures benefit from multi-channel conditional input, with a slight improvement in performance comparing cGAN$_{1}$ to cGAN$_{3}$, and a significant improvement when comparing cDDPM$_{1}$ to cDDPM$_{3}$. The best SIMOS score was achieved by cDDPM$_{3}$.

\section{Discussion and Conclusion}

We aimed to conduct a fair and unbiased comparison of the cGAN and cDDPM architectures for MRI-to-CT translation. However, some challenges are introduced by the specific implementations used. In particular, the computationally intensive nature of the cDDPM models meant that the approaches were difficult to compare under similar compute budgets. This could have been mitigated by using another noise schedule, such as a cosine noise schedule, which, according to Nichol and Dhariwal \cite{nichol2021improvedBETTER_DDPM}, introduces a 'negligible' difference in quality while lowering sampling time.

\begin{figure}[h]
    \centering
    \begin{subfigure}{0.2\textwidth}
        \centering
        \includegraphics[width=\linewidth]{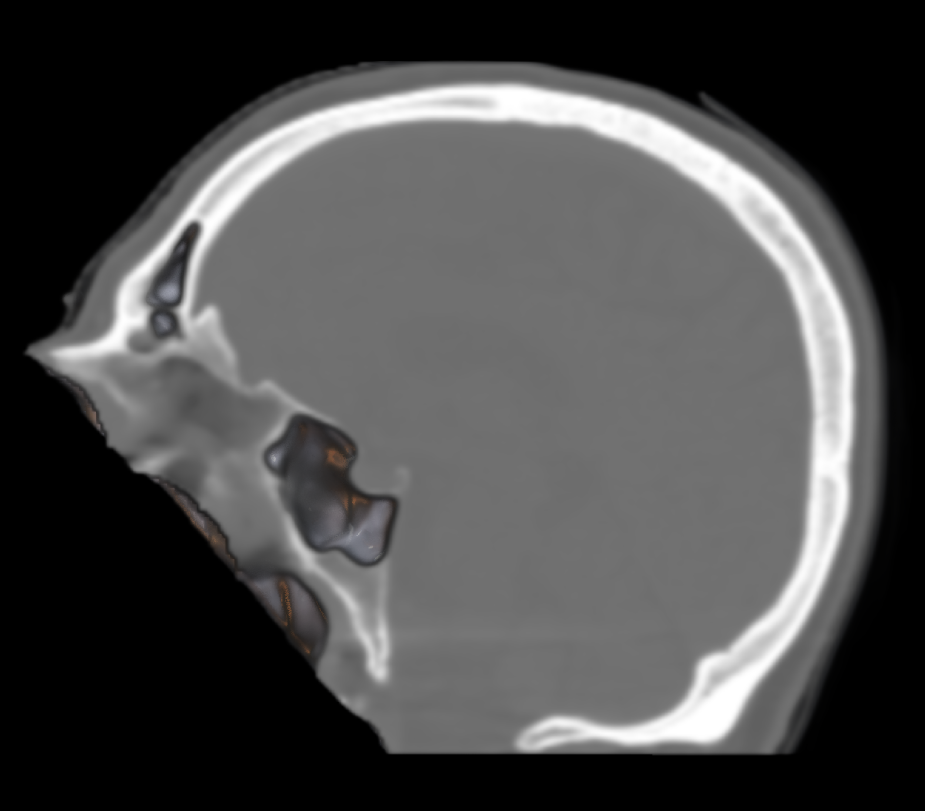} 
        \begin{minipage}{0.8\linewidth} 
            \caption{GT}
        \end{minipage}
    \end{subfigure}%
    \hfill 
    \begin{subfigure}{0.2\textwidth}
        \centering
        \includegraphics[width=\linewidth]{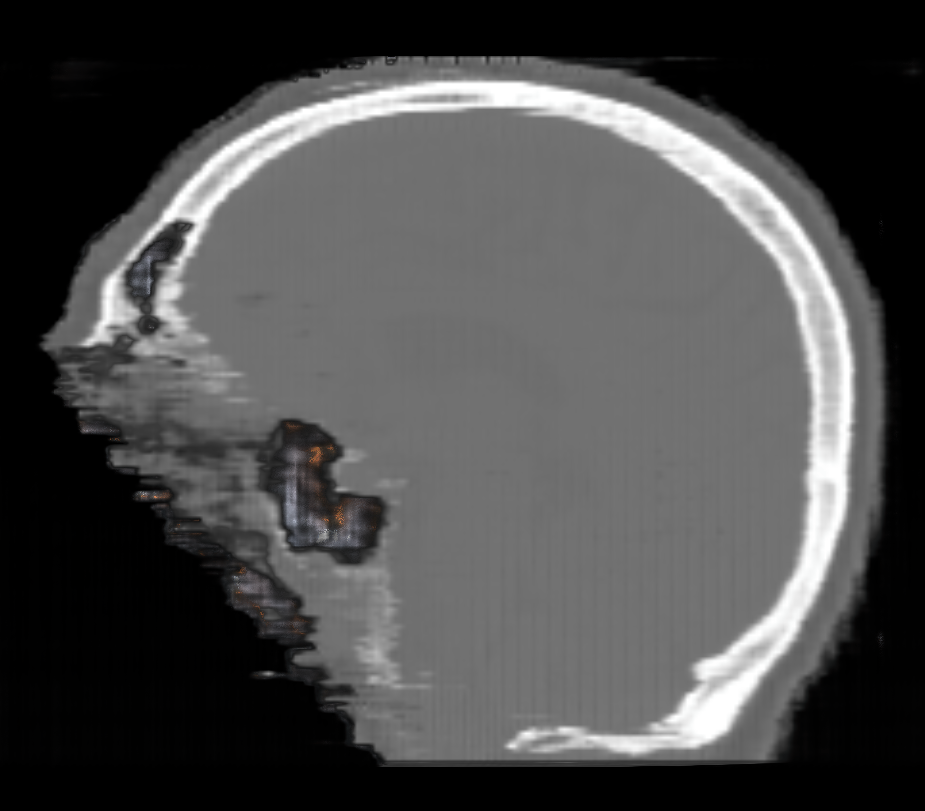} 
        \begin{minipage}{0.8\linewidth} 
            \caption{cGAN$_1$}
        \end{minipage}
    \end{subfigure}%
    \hfill 
    \begin{subfigure}{0.2\textwidth}
        \centering
        \includegraphics[width=\linewidth]{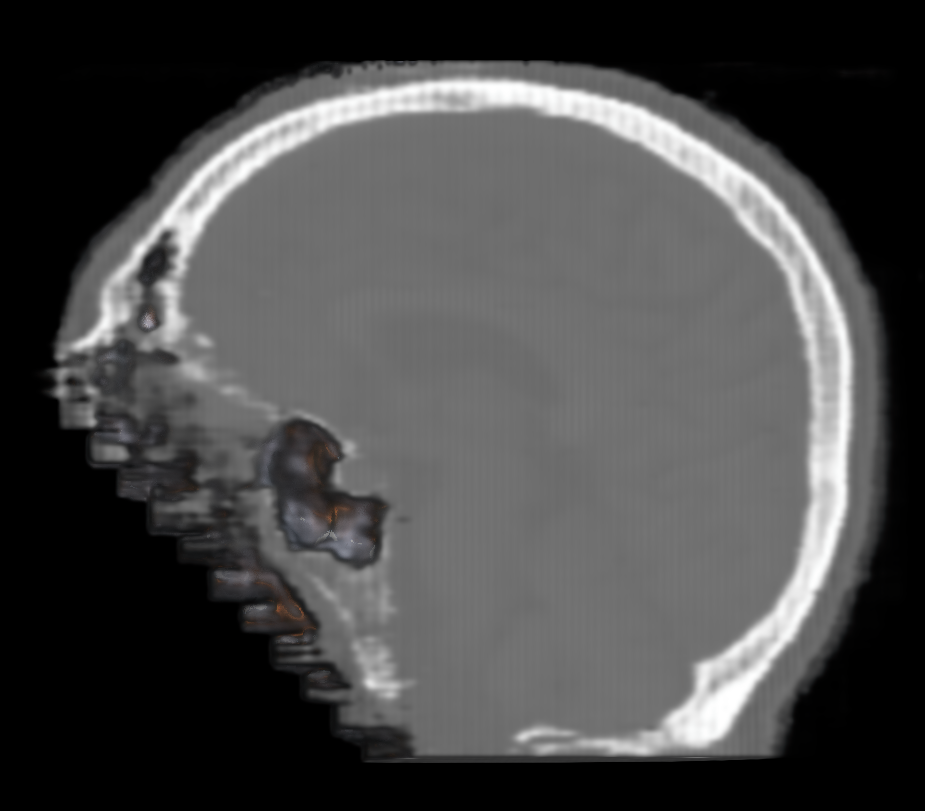}
        \begin{minipage}{0.8\linewidth}
            \caption{cGAN$_3$}            
        \end{minipage}
    \end{subfigure}%
    \hfill
    \hfill 
    \begin{subfigure}{0.2\textwidth}
        \centering
        \includegraphics[width=\linewidth]{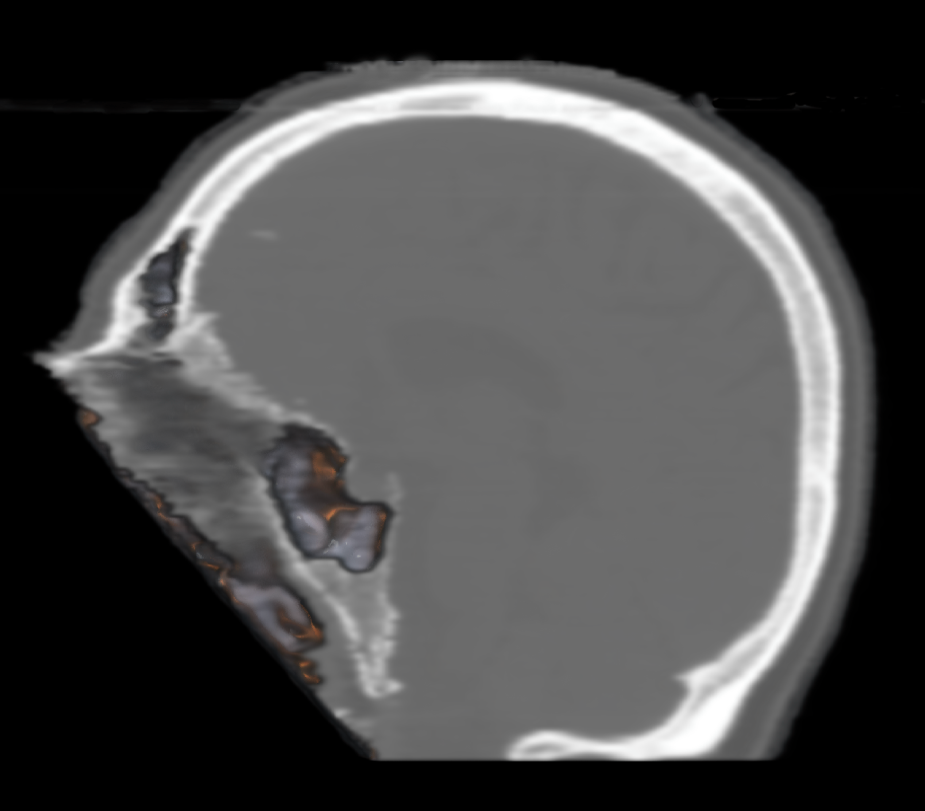}
        \begin{minipage}{0.8\linewidth}
            \caption{cDDPM$_1$}        
        \end{minipage}
    \end{subfigure}%
    \hfill    
    \begin{subfigure}{0.2\textwidth}
        \centering
        \includegraphics[width=\linewidth]{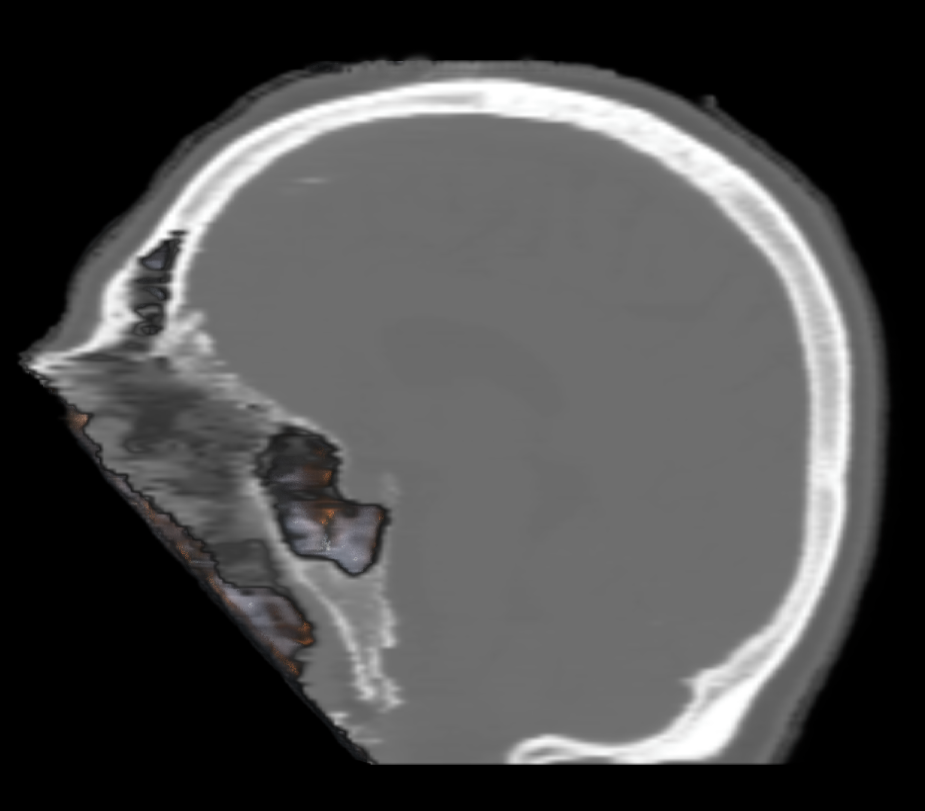}
        \begin{minipage}{0.8\linewidth}
            \caption{cDDPM$_3$}            
        \end{minipage}
    \end{subfigure}
    \caption{Brain samples.}
    \label{fig:visual_inspection_brain}
\end{figure}

\begin{figure}[h]
    \centering
    \begin{subfigure}{0.2\textwidth}
        \centering
        \includegraphics[width=\linewidth]{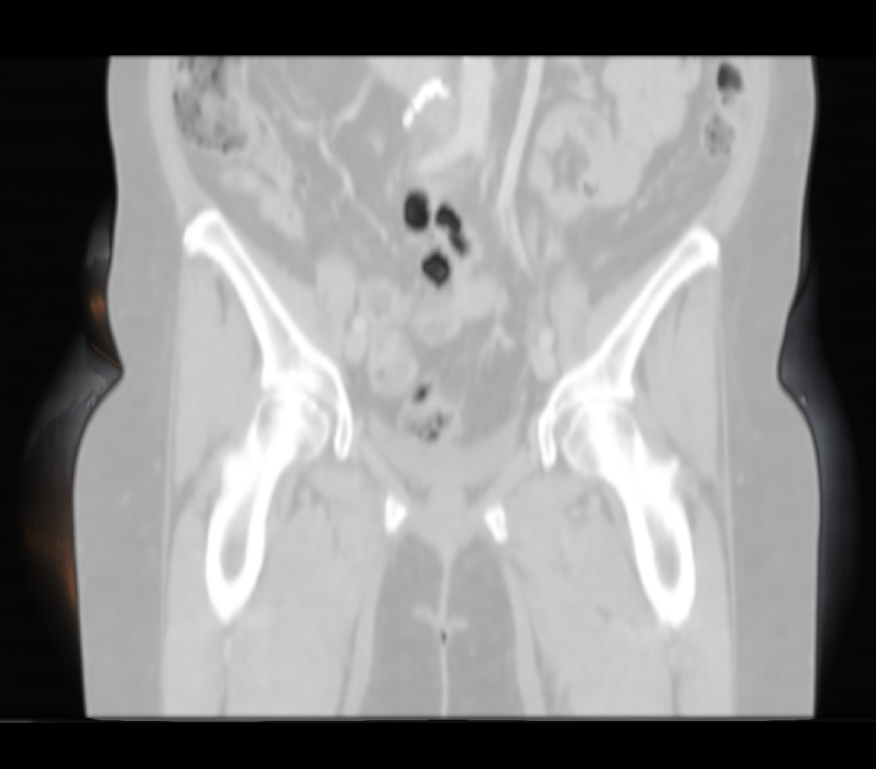} 
        \begin{minipage}{0.8\linewidth} 
            \caption{GT}
        \end{minipage}
    \end{subfigure}%
    \hfill 
    \begin{subfigure}{0.2\textwidth}
        \centering
        \includegraphics[width=\linewidth]{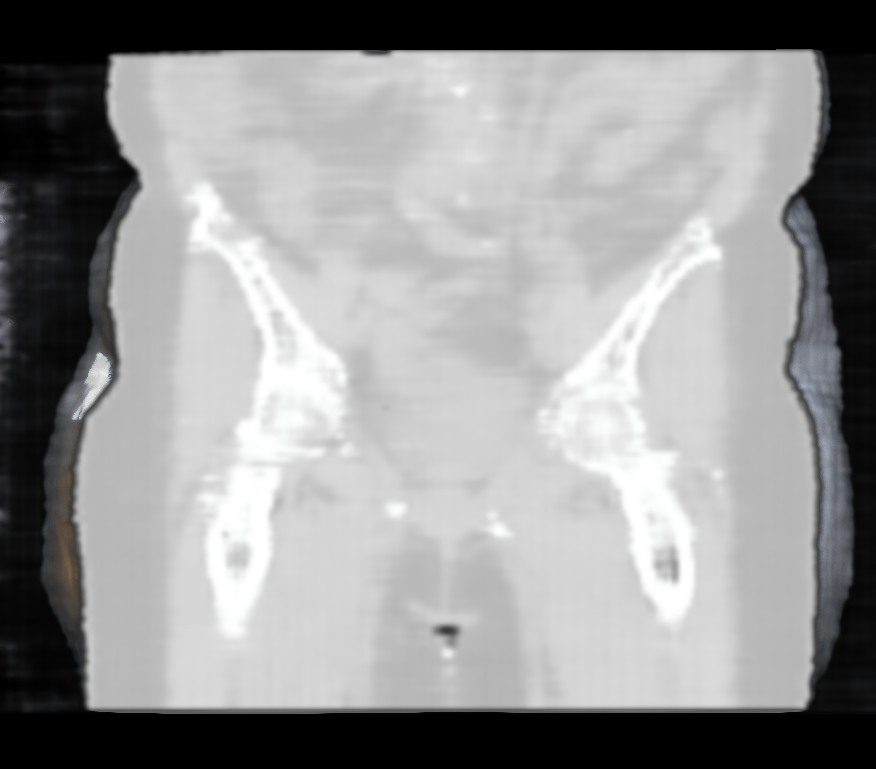} 
        \begin{minipage}{0.8\linewidth} 
            \caption{cGAN$_1$}
        \end{minipage}
    \end{subfigure}%
    \hfill 
    \begin{subfigure}{0.2\textwidth}
        \centering
        \includegraphics[width=\linewidth]{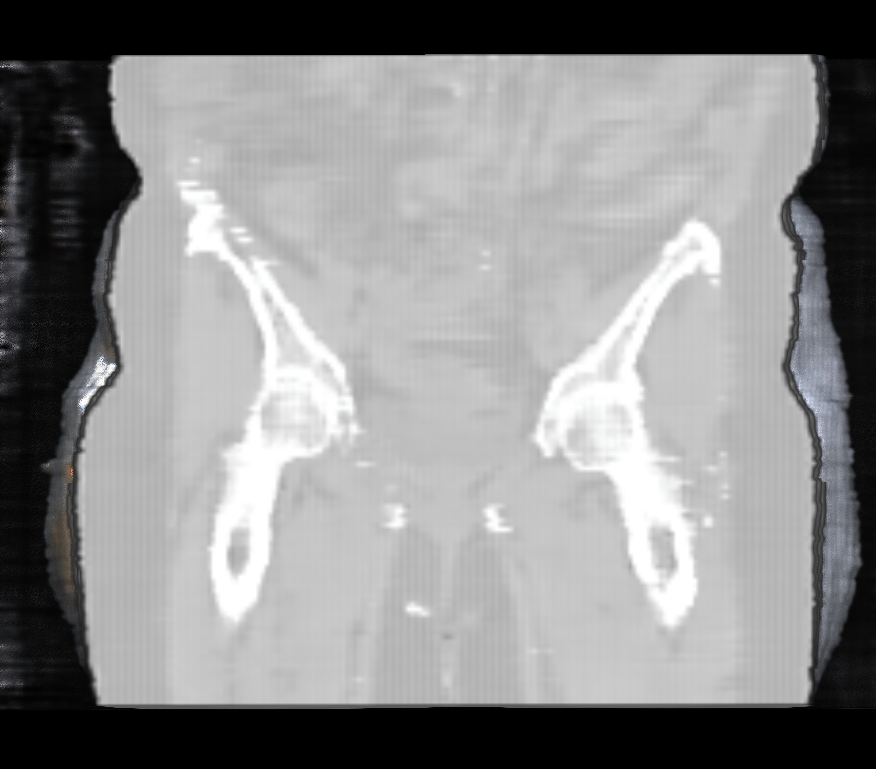}
        \begin{minipage}{0.8\linewidth}
            \caption{cGAN$_3$}            
        \end{minipage}
    \end{subfigure}%
    \hfill
    \hfill 
    \begin{subfigure}{0.2\textwidth}
        \centering
        \includegraphics[width=\linewidth]{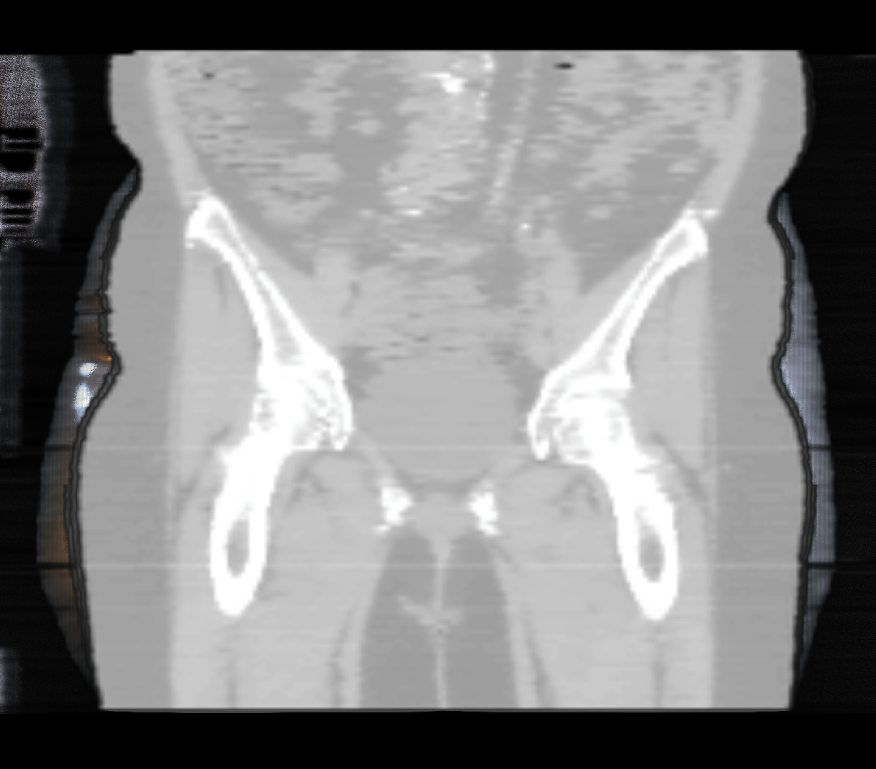}
        \begin{minipage}{0.8\linewidth}
            \caption{cDDPM$_1$}        
        \end{minipage}
    \end{subfigure}%
    \hfill    
    \begin{subfigure}{0.2\textwidth}
        \centering
        \includegraphics[width=\linewidth]{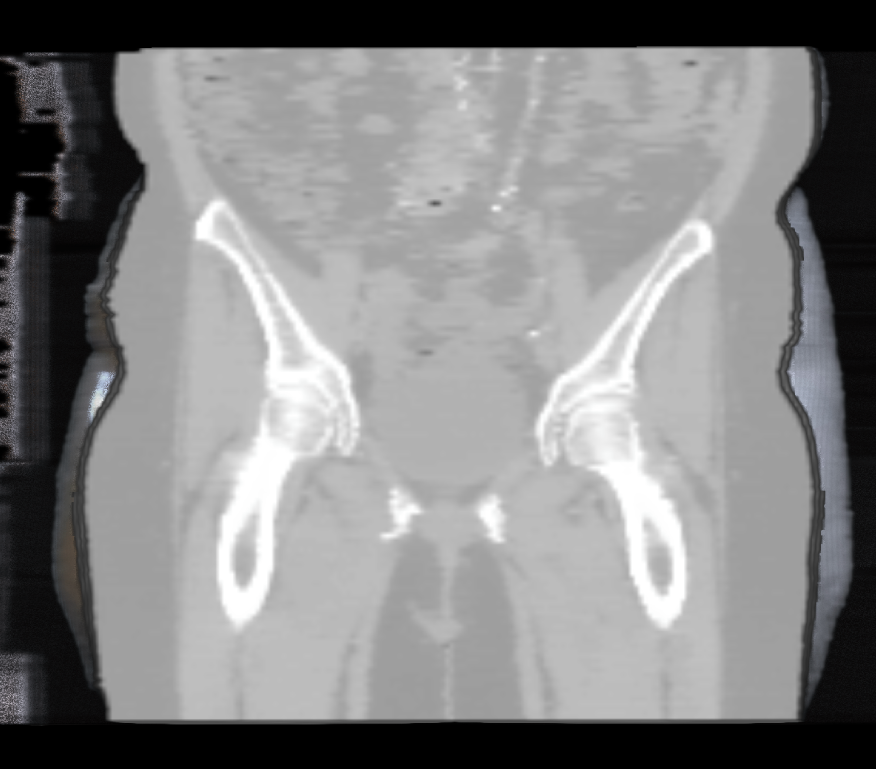}
        \begin{minipage}{0.8\linewidth}
            \caption{cDDPM$_3$}            
        \end{minipage}
    \end{subfigure}
    \caption{Samples of the pelvic region.}
    \label{fig:visual_inspection_pelvic}
\end{figure}

Visual inspection of the sCT when resampled into full scans reveals that all models have a tendency to blur soft tissue regions, which is more pronounced in the cGAN models than the cDDPM models. The cDDPM-based models manage to generate sCTs with less discontinuity between slices and seem to produce more faithful results than cGAN-based models (see Fig. \ref{fig:visual_inspection_brain}, \ref{fig:visual_inspection_pelvic}).


To evaluate applications in radiotherapy, our evaluation protocol could have been extended with a comparison of a treatment plan based on the sCT and the ground truth CT. Utilizing this as a metric would test the generative nets' ability to correctly synthesise the radiodensity of tissue in 3D. This would be highly relevant for many practical medical applications of CT images.

\subsubsection{Impact of multi-channel conditionals.}
Results indicate that multi-channel input improves the quality of the generated sCT. The effect is most pronounced for the cDDPM models, where cDDPM$_3$ outperforms cDDPM$_1$ on four metrics while using approximately the same computation time. In the cGAN models, the effect is less obvious, though we see a significantly better SIMOS score for cGAN$_3$ than for cGAN$_1$. Generally, the multi-channel models score lower SIMOS values, indicating that providing the model with more spatial information results in more continuity across sampled slices. The most significant improvement from multi-channel input is detected in the cDDPM-based model.

\subsubsection{SynthRAD2023.}
The SynthRAD2023 Grand Challenge \cite{challengeDATA} aimed to generate sCT from MRI. The competition is finalized, and the scoreboard is accessible at \cite{synthRad}. To identify a solution comparable to ours, we surveyed the top five entries. A fair comparison is only possible if the models are trained on the same data and the preprocessing pipelines are similar.

The fourth place submitted by Alain-Beaudoin et al. \cite{alain2023pairedELEKTRA} qualified under these criteria. They decreased the Hounsfield range to $[-1000; 2200]$ compared to our $[-1000; 2000]$ and normalized the MRI locally by a percentile-determined range. They scored a PSNR of $28.64 \pm 1.77$, and an SSIM of $0.872 \pm 0.032$ in the validation phase \cite{alain2023pairedELEKTRA}. Our cDDPM-based models achieve a better or equal SSIM score, but the SSIM of our cGAN-based models are lower. This could mean that the cDDPM models are better suited for maintaining structures in the image, this is backed by higher IoU for these models. Compared to our results the PSNR of the model presented by Alain-Beaudoin et al. \cite{alain2023pairedELEKTRA} is better.

\subsubsection{Summary.}
Both approaches are viable for MRI-to-CT translation. The cDDPM architecture is more suitable for the task, as it achieves better scores in the SSIM, PSNR, FID, SIMOS and segmentation. Though the computational cost is considerably higher for this architecture, this difference could be decreased by sampling with another noise schedule. Visual inspection reveals satisfactory results for both architectures, but the cDDPM does perform better in this aspect of the evaluation as well.

Multi-channel conditional input affected the cDDPM architecture more than the cGAN, but the overall result of providing this additional information was beneficial.

\bibliographystyle{splncs04}
\bibliography{litList}

\end{document}